\documentclass[twocolumn,showpacs,preprintnumbers,amsmath,amssymb]{revtex4}
\usepackage{graphicx}% Include figure files
\usepackage{dcolumn}% Align table columns on decimal point
\usepackage{bm}% bold math

\begin{document}

\preprint{}

\title{Control of light speed: From slow light to superluminal light}


\author{Qun-Feng Chen}
%\email[]{qfchen@mail.ustc.edu.cn}
\author{Yong-Sheng Zhang}
%\email[]{yshzhang@ustc.edu.cn}
\author{Bao-Sen Shi}
\author{Guang-Can Guo}
%\thanks{}
\affiliation{Key Laboratory of Quantum Information, University of Science and
Technology of China, Hefei, 230026, Pepple's Republic of China}


\date{\today}

\begin{abstract}
  A scheme for controlling light speed from slower-than-$c$ to faster-than-$c$ in an
  atomic system is presented in this paper. The scheme is based on far detuning
  Raman effect. Two far detuning coupling fields with small frequency
  difference will produce two absorptive peaks for the probe field in a
  $\Lambda$ structure, and an optical pump between the two ground states can
  change the absorptive peaks into enhanced peaks, which makes the normal
  dispersion between the two peaks change into anomalous dispersion, so the
  probe field can change from slow light to superluminal propagation.
\end{abstract}

%\keywords{}

\maketitle

Control of light speed has attracted much attention in the past years.  The
slow light has been demonstrated in both atomic vapor
\cite{PhysRevLett.74.2447, PhysRevA.53.R27, NATURE.397.594,
PhysRevLett.82.5229, PhysRevLett.83.1767} and solid systems
\cite{PhysRevLett.88.023602,PhysRevLett.90.113903}, most of them
\cite{PhysRevLett.74.2447, PhysRevA.53.R27, NATURE.397.594,
PhysRevLett.82.5229, PhysRevLett.83.1767, PhysRevLett.88.023602} are based on
the electromagnetically induced transparency
(EIT)\cite{HFI90,Fleischhauer:2005}. The first superluminal light propagation
without large absorption or reshaping using far detuning Raman enhancement was reported in
Ref.~\cite{NATURE.406.277}.  Recently,
experiments in solid systems show that both superluminal and subluminal propagations can be
obtained in the same system \cite{MatthewS.Bigelow07112003, JPCM:R1321,
longhi:056614, OE.14.6201}. In addition, some theoretical proposals
\cite{PhysRevA.63.043818, PhysRevA.64.053809, PhysRevA.68.043818} show the
possibility of changing light propagation from subluminal to superluminal in an
atomic system. However, there are some drawbacks in these schemes, in which
either coupling of the dipole transition forbidden
levels\cite{PhysRevA.63.043818,PhysRevA.64.053809} or some very special
levels structure is required\cite{PhysRevA.68.043818}, which makes these schemes
difficult to be realized. Furthermore, they do not show a good relation between the
controlling and the dispersion of the system either.

\begin{figure}[b]
  \begin{center}
	\includegraphics[width=0.35\textwidth]{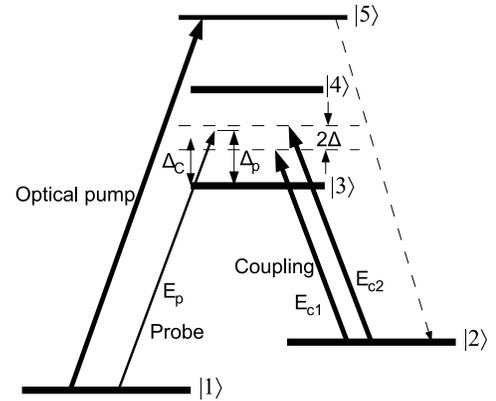}
  \end{center}
  \caption{Atomic level diagram. $E_{c1}$ and $E_{c2}$ are the two coupling
  fields, which have equal intensity and a small frequency difference of
  2$\Delta$. The center frequency of the two coupling fields has a detuning of $\Delta_c$
  from $\left|2\right>\to\left|3\right>$ transition.  $E_p$ is the probe
  field, which is detuning from $\left|2\right>\to\left|3\right>$ transition by
  $\Delta_p$. Level $\left|5\right>$ is used for optical pump, the pumping rate
  is $R_{\rm op}$.}
  \label{fig1}
\end{figure}

In this paper we give a scheme which can be used to change the group velocity
of probe field from slower-than-$c$ to faster-than-$c$ continuously by controlling
the strength of an optical pump. The dispersion of the system can be changed
from positive to negative with the increment of the pumping rate. At
the same time the probe field will not encounter large absorption or reshaping
in large range of dispersion change. This scheme is based on far detuning Raman effect, so it is easy to
be realized in many atomic systems, e.g. in cold or hot rubidium system.
The atomic level structure used in this scheme is a five-level structure,
as shown in Fig.~\ref{fig1}.
Four of the five levels (levels $\left|1\right>$ to $\left|4\right>$) are used
to create a far detuning symmetric $\Lambda$ structure. The electronic dipole
moments between the ground states (levels $\left|1\right>$ and $\left|2\right>$)
and the excited states (levels $\left|3\right>$ and $\left|4\right>$) satisfy
the following relation: three of the dipole moments have the same phase while
the phase of the forth is opposite to them. This is the requirement to get far
detuning Raman absorptive peak when the coupling field lies at the center of
the $\left|2\right>\to\left|3\right>$ and $\left|2\right>\to\left|4\right>$
transitions. This requirement can be fulfilled in many systems, for example
rubidium 87, sodium and cesium systems\cite{DLineData}. In this scheme two
coupling fields ($E_{c1}$ and $E_{c2}$) with a small frequency difference of
$2\Delta$ are used. The center frequency of them is set to the center of the
$\left|2\right>\to\left|3\right>$ and $\left|2\right>\to\left|4\right>$
transitions. The probe field $E_p$ is detuning from the
$\left|1\right>\to\left|3\right>$ transition by $\Delta_p$.  These three fields
form a far detuning $\Lambda$ structure. The two detuning coupling fields
produce two Raman absorptive peaks for the probe field. Therefore the probe
field will encounter normal dispersion and the group velocity will be slowed
down when its frequency lies between the two Raman absorptive peaks. In order
to change the group velocity of the probe field, an optical pump from
$\left|1\right>$ to $\left|2\right>$ is applied to the system (using the level
$\left|5\right>$). As the strength of the optical pump is increased gradually,
the Raman absorptive peaks will become lower and finally change into enhanced
peaks, at the same time the full width at half maximum (FWHM) of the two Raman
peaks are almost unchanged\cite{PRA.73.053804}. Therefore the dispersion
between the two Raman peaks changes from positive to negative, and the group
velocity of the probe field changes from slower-than-$c$ to faster-than-$c$.

We show our main results based on numerical calculation. Using the
definition of group velocity $v_g\equiv{d \omega/dk}$ and the relation between
index of refraction $n$ and the susceptibility $\chi$: $n=\sqrt{1+\chi}$, and
ignoring the absorption or enhancement we get\\
\begin{equation}
  \frac{1}{v_g}=\frac{1} {c} \left[\sqrt{1+{\rm
  Re}\chi}+\frac{\omega}{2\sqrt{1+{\rm
  Re}\chi}}\frac{d({\rm Re}\chi)}{d\omega}\right].
	\label{vg}
\end{equation}
The index of group velocity is defined as $n_g\equiv{c}/{v_g}$.
Equation~(\ref{vg}) indicates that  $(n_g-1)\propto d({\rm Re}\chi)/
d\omega$ when ${\rm Re}\chi\ll1$, so in the following we concentrate on $d[{\rm Re}\chi(\omega_p)]/
d\omega_p$, where $\chi(\omega_p)$ is the susceptibility of the system for the
probe field. We solve the master equation for the atomic density operator to
get the $\chi(\omega_p)$.

Ignoring the optical pump, the effective Hamiltonian of this system can be
written as\cite{Fleischhauer:2005}:
\begin{widetext}
\begin{equation}
  H=-\frac{\hbar}{2}\left(\begin{array}{cccc}
	0 & 0 & \Omega_{p3} & \Omega_{p4}\\
	0 & 2(\Delta_p-\Delta_c) & \Omega_{c3}(e^{i \Delta t}+e^{-i \Delta t}) &
	\Omega_{c4}(e^{i \Delta t}+e^{-i \Delta t}) \\
	\Omega_{p3} & \Omega_{c3}(e^{i \Delta t}+e^{-i \Delta t}) & 2\Delta_p & 0 \\
	\Omega_{p4} & \Omega_{c4}(e^{i \Delta t}+e^{-i \Delta t}) & 0 &
	2(\Delta_p-\omega_{43})
  \end{array}\right)\;,
  \label{Hamiltonian}
\end{equation}
\end{widetext}
where $\Delta_p=\omega_p-\omega_{31}$,
$\Delta_c=(\omega_{c1}+\omega_{c2})/2-\omega_{32}$,
$\Delta=(\omega_{c2}-\omega_{c1})/2$, $\omega_{ij}$ is the frequency difference
between levels $\left|i\right>$ and $\left|j\right>$,
$\Omega_{pi}=\mu_{i1}E_p/\hbar$ and $\Omega_{ci}=\mu_{i2}E_c/\hbar$ are the
Rabi frequencies of the fields with the corresponding transitions, and $\mu_{ij}$
is the transition electronic dipole moments of the
$\left|i\right>\to\left|j\right>$ transition. Here we suppose all the
$\Omega_{pi}$ and $\Omega_{ci}$ are real. The master equation for the atomic
density operator can be written as \cite{Fleischhauer:2005}:
\begin{eqnarray}
  \frac{d\rho}{dt}&=&\frac{1}{i\hbar}[H,\rho]+\frac{\Gamma_{31}}{
  2}[2\hat\sigma_{13}\rho\hat\sigma_{31}-\hat\sigma_{33}\rho-\rho\hat\sigma_{33}]\nonumber\\
  &&+\frac{\Gamma_{32}}{2}[2\hat\sigma_{23}\rho\hat\sigma_{32}-\hat\sigma_{33}\rho-\rho\hat\sigma_{33}]
  \nonumber \\
  &&+\frac{\Gamma_{41}}{
  2}[2\hat\sigma_{14}\rho\hat\sigma_{41}-\hat\sigma_{44}\rho-\rho\hat\sigma_{44}]\nonumber\\
  &&+\frac{\Gamma_{42}}{
  2}[2\hat\sigma_{24}\rho\hat\sigma_{42}-\hat\sigma_{44}\rho-\rho\hat\sigma_{44}]\nonumber\\
  &&+\frac{\gamma_{3\rm deph}}{2}[2\hat\sigma_{33}\rho\hat\sigma_{33}-\hat\sigma_{33}\rho-\rho\hat\sigma_{33}]
  \nonumber \\
  &&+\frac{\gamma_{4\rm deph}}{
  2}[2\hat\sigma_{44}\rho\hat\sigma_{44}-\hat\sigma_{44}\rho-\rho\hat\sigma_{44}]
  \nonumber \\
  &&+\frac{\gamma_{\rm2deph}}{
  2}[2\hat\sigma_{22}\rho\hat\sigma_{22}-\hat\sigma_{22}\rho-\rho\hat\sigma_{22}]\nonumber\\
  &&-R_{\rm op}\rho_{11}(\hat\sigma_{11}-\hat\sigma_{22})\;,
  \label{master}
\end{eqnarray}
where the terms from the second to the fifth on the right-hand side describe
spontaneous emission from excited states to the ground states, the
$\Gamma_{ij}$ is the spontaneous emission rate from $\left|i\right>$ to $\left|j\right>$, the $\gamma_{i\rm deph}$ is the dephasing rate of the state
$\left|i\right>$, and the last term represents the optical pump effect,
$R_{\rm op}$ is the pump rate. When the
optical pump is a single frequency laser and only one excited level is used for optical pump, the 
relation between the pump rate and the Rabi frequency of the optical pump field
$\Omega_{\rm op}$ derived from a three-level structure is
\begin{equation}
  R_{\rm
  op}=\frac{\Gamma_{52}}{\displaystyle\frac{\Gamma_{5}(\gamma_{51}^{2}+4\Delta_{op}^{2})}{\gamma_{51}\Omega_{\rm
  op}^{2}}+1},
  \label{ropeq}
\end{equation}
where $\Gamma_{5}=\Gamma_{51}+\Gamma_{52}$ is the total spontaneous emission
rate out of $|5\rangle$, $\gamma_{51}=\Gamma_{51}+\gamma_{5\rm deph}$,
$\Gamma_{ij}$ and $\gamma_{5\rm deph}$ are defined the same as the above
definitions,
$\Delta_{\rm op}$ is the detuning of the optical pump field. Equation
(\ref{ropeq}) gives the upper bound of the optical pump rate: $R_{\rm op}<\Gamma_{52}$.

In order to simplify the calculation, we suppose 
$\mu_{32}=\mu_{42}=\mu_{31}=-\mu_{41}$, so that
$\Gamma_{32}=\Gamma_{42}=\Gamma_{31}=\Gamma_{41}$,
$\Omega_{c3}=\Omega_{c4}=\Omega_c$ and $\Omega_{p3}=-\Omega_{p4}=\Omega_{p}$.
The expression of the susceptibility for the probe field can be written as:
\begin{equation}
\chi(\omega_p)=k(\rho_{31}/\Omega_{p3} + \rho_{41}/\Omega_{p4})\;,
\end{equation}
where $k=\varrho|\mu_{31}|^2/(\epsilon_0\hbar)$ is a constant related to
atomic density $\varrho$ and transition electronic dipole moment
$\mu_{31}(\mu_{41})$ \cite{Fleischhauer:2005}. In the following calculation we
treat the $\rho_{31}/\Omega_{p3} + \rho_{41}/\Omega_{p4}$ as $\chi(\omega_p)$,
since there is only a constant scale difference between them. We write
$\chi(\omega_p)$ as $\chi$ in the follows. For convenience, we define the total
spontaneous emission rate out of state $\left|3\right>$ as
$\Gamma_{3}=\Gamma_{31}+\Gamma_{32}$. The dephasing decays from the excited
levels are ignored, since they are much smaller than the spontaneous emissions
in normal system. We set $\gamma_{2\rm deph}=0.01\Gamma_3$,
$\omega_{43}=140\Gamma_3$, which is got from the energy split of $5P_{1/2}$ of
$^{87}$Rb, as the system parameters, and set $\Delta_c=\omega_{43}/2$ and
$\Omega_p=0.01\Gamma_3$ as the numerical conditions.

\begin{figure}
  \begin{center}
	\includegraphics[width=0.48\textwidth]{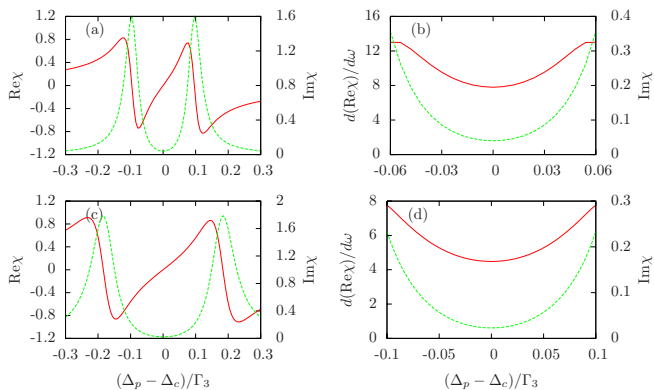}
  \end{center}
  \caption{(Color online) $\chi$ and $d\chi/d\omega$ with different values of
  coupling fields strength and $\Delta$ when $R_{\rm op}=0$. 
  Left figures show the $\chi$ versus two photon detuning, in
  which the red solid line and green dashed line represent the real and
  imaginary parts of $\chi$, respectively. The right figures show the real
  part of $d\chi/d\omega$ (red solid line) and the imaginary part of $\chi$
  (green dashed line)
  between the two Raman peaks.
  (a),(b) $\Omega_c=20\Gamma_3$, $\Delta=0.1\Gamma_3$; (c),(d)
  $\Omega_c=30\Gamma_3$, $\Delta=0.2\Gamma_3$.}
  \label{fig2}
\end{figure}

Firstly, the master equation is solved with the absence
of the optical pump. The $\chi$ and $d\chi/d\omega$ versus two photon detuning
for different values of coupling fields strength are given in Fig.~\ref{fig2}.
The figure shows that the two far detuning coupling fields generate two Raman
absorptive peaks, and there is a positive
dispersion at the center of these two peaks, which will slow down the probe field. The $d({\rm Re}\chi)/d\omega$
and Im$\chi$ between the two peaks are shown at the right side of Fig.~\ref{fig2},
the $d({\rm Re}\chi)/d\omega$ and Im$\chi$ determine the group velocity and
absorption of the probe field.  The figures show that at the center of the two
peaks dispersion is much larger than absorption.  The figures also show that
the bandwidth of the system is mainly determined by $\Delta$. Larger $\Delta$
leads to larger bandwidth but smaller dispersion at the same coupling strength.

\begin{figure}
	\begin{center}
		\includegraphics[width=0.48\textwidth]{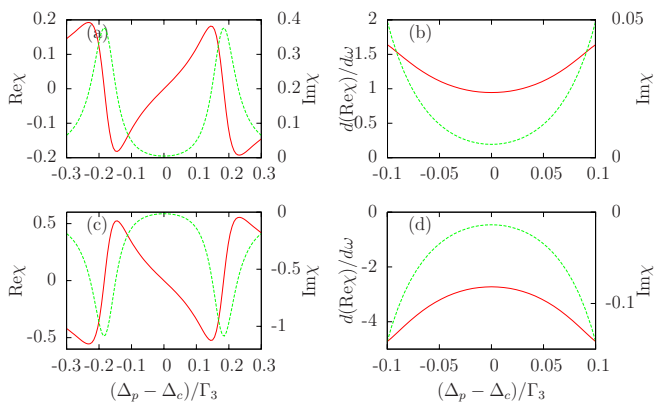}
	\end{center}
	\caption{(Color online) $\chi$ and $d\chi/d\omega$ versus two photon
	detuning with different optical
	pump rates, when $\Omega_c=30\Gamma_3$ and $\Delta=0.2\Gamma_3$. The
	figures layout are the same as Fig.~\ref{fig2}.(a),(b) $R_{\rm op}=0.06$;
	(c),(d) $R_{\rm op}=0.4\Gamma_{3}$.}
	\label{fig3}
\end{figure}

When an optical pump from $\left|1\right>$ to $\left|2\right>$ is applied to
the system, the Raman peaks change obviously. The $\chi$ and
$d({\rm Re}\chi)/d\omega$ versus two photon detuning with the optical pump
rates of 0.06$\Gamma_3$ and 0.4$\Gamma_3$ when $\Omega_c=30\Gamma_3$ and
$\Delta=0.2\Gamma_3$ are shown in Fig.~\ref{fig3}. Figures ~\ref{fig3}(a) and \ref{fig3}(b) show
that when an optical pump is added to the system, the dispersion and absorption
become much smaller compared with Figs.~\ref{fig2}(c) and \ref{fig2}(d), while
the shape of $\chi$ versus two photon detuning almost unchanged.
Figures~\ref{fig3}(c) and \ref{fig3}(d) show the result of a sufficient large optical pump
added to the system. The two Raman absorptive peaks change into enhanced
peaks, the normal dispersion changes into anomalous dispersion, while the
FWHM of the Raman peaks and the ratio of the dispersion to the
absorption(enhancement) of the system remain almost unchanged. The experimental
demonstration of the control of the Raman peak using an optical pump has been
done in our previous work \cite{PRA.73.053804} in room temperature paraffin
coated rubidium 87 cell.

\begin{figure}
  \begin{center}
	\includegraphics[width=0.45\textwidth]{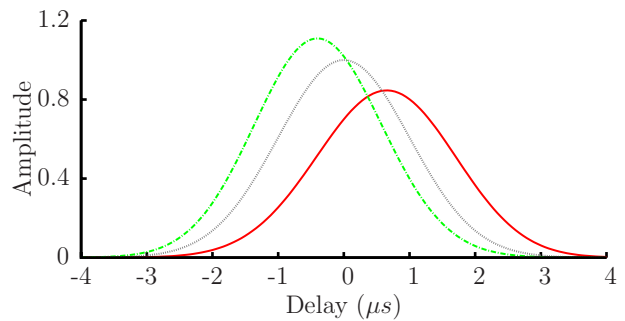}
  \end{center}
  \caption{(Color online)
  Simulation of a 1 $\mu s$ Gaussian wave packet transmitting through 1
  mm cold $^{87}$Rb cloud with atomic density of $5\times10^{11}/$cm$^3$.
  The frequency difference of the coupling fields is 2$\Delta=0.4\Gamma_3$.
  The Rabi frequencies of the coupling fields are
  $\Omega_c=30\Gamma_3$. The red solid line is corresponding to the optical pump
  rate of $R_{\rm op}=0$, and $n_g=1.9\times10^5$. The green dashed line is
  corresponding to the optical pump rate of $R_{\rm op}=0.4\Gamma_3$, and
  $n_g=-1.1\times10^5$.  The gray dot line is the reference. 
  }
  \label{fig4}
\end{figure}

\begin{figure}
  \begin{center}
	\includegraphics[width=0.45\textwidth]{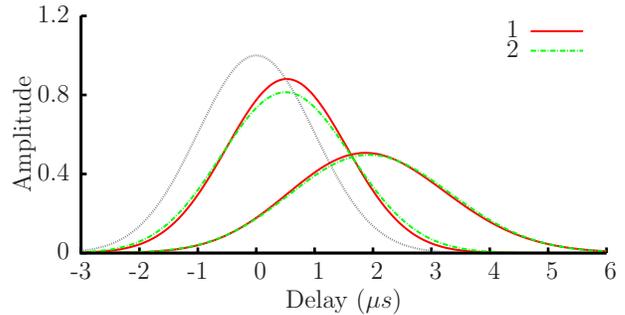}
  \end{center}
  \caption{(Color online)
  Comparing the slow light effect with the EIT scheme. Line~1 is the slow light
  effect of EIT with different coupling strength, wave packets from right to
  left are corresponding to $\Omega_c= 0.5\Gamma_3$ and $1\Gamma_3$, and
  $n_g=5.9\times10^5$ and $1.6\times10^5$. Line~2 is the slow light effect with
  different optical pump rate when $\Omega_c=55\Gamma_3$, wave packets from
  right to left are corresponding to $R_{\rm op}=0$, $0.17\Gamma_3$ and
  $n_g=6.0\times10^5$ and $1.6\times10^5$.  The gray dot line is the reference.
  }
  \label{fig5}
\end{figure}
To show the slow and fast light effect of this two-coupling scheme, we
simulate a 1
$\mu s$ Gaussian probe field (FWHM$\:\approx2.35\;\mu s$) transmitting
through a 1-mm cold $^{87}$Rb cloud with the
atomic density of $5\times10^{11}/$cm$^3$. The $D1$
transitions used to create the
$\Lambda$ system and the $D2$ transition for optical pump. The frequency difference of the two
coupling is set to $0.4\Gamma_3$.
In the simulation we suppose that the probe field is weak and almost do not
influence the system.
The slow light and fast light effects with the Rabi frequencies of the coupling
fields of $30\Gamma_3$ are shown in Fig.~\ref{fig4}. This figure shows clearly
that when the optical pump is absent the probe field gets a positive group delay, while
when certain optical pump is added, the probe field gets a negative group delay.
This figure also shows that the probe field encounters little reshape,
stretched by a factor of about 1.03 with the positive group delay,
while narrowed by a factor of 0.95 with the negative group delay. In our
simulation
the probe pulse is a 1 $\mu s$ Gaussian wave packet, which corresponds to
$\Delta\omega=1\times10^{6}$ rad/$s$ and FWHM$\:\approx2.35\times10^{6}$ rad/$s$. While the FWHM of the 
transmission width of the system is about $6.6\times10^{6}$
rad/$s$.
The line width of the probe field is not far smaller than the
transmission width, this causes
the stretch and narrowing of the wave packet with positive and negative
group delay.  There are also some internal reasons which make the wave
packet narrowed with the negative group delay.  A system can not make
a wave packet pass through before the arrival of this wave packet. The
speed of the starting point of a wave packet should be slower than or
equal to $c$. Therefore if the wave packet gets a negative group delay, it should be compressed in the time domain.
 
In order to examine the performance of this two-coupling scheme,
a comparison between this two-coupling scheme and the normal EIT scheme in the
slow light region is given. The EIT scheme is the best slow light scheme in the atomic
system to our best knowledge.
The system parameters used are the same as the above simulation.
The results are shown in Fig.~\ref{fig5}.
We compare the two schemes with almost the same delay occurred. In
Fig.~\ref{fig5}, line~1 shows the results of the EIT scheme with different
coupling strengths and line~2 shows the results of the two-coupling scheme with
different pump rates. The figure shows that when the optical pump is absent the
two-coupling scheme almost has the same effect with the EIT scheme, the same
delay causes the same stretch, as shown in the right wave packets of
Fig.~\ref{fig5}. This demonstrates that in the slow light region this two-coupling
scheme is comparable with the EIT scheme.
In the EIT scheme the coupling power is increased to make the probe pulse get less
delay, and as the increment of the coupling power the transmission width for
the probe field will
get larger, while if we use the optical pump in the two-coupling scheme to make
the probe field get less delay, the transmission width do not change.
Therefore in this case the wave packet delayed by the two-coupling scheme will
get a larger stretch than the EIT scheme, as shown in the left wave packets of
Fig.~\ref{fig5}.

\begin{figure}
  \begin{center}
	\includegraphics[width=0.4\textwidth]{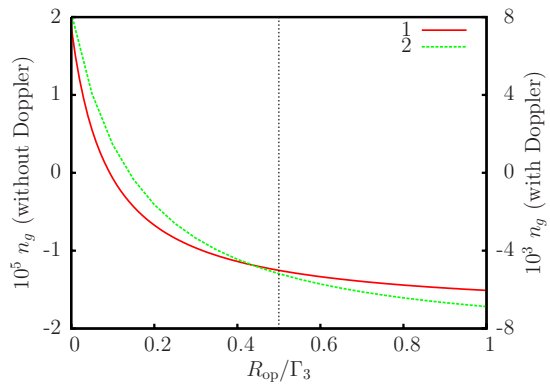}
  \end{center}
  \caption{(Color online) The relation between $n_g$ and 
  the optical pump rate when
  $\Omega_c=30\Gamma_3$, $\Delta=0.2\Gamma_3$, using the same system
  parameters as Fig.~\ref{fig4}. Line 1: without Doppler broadening, line 2:  
 with Doppler broadening. The
  Doppler broadening is calculated using the data of $^{87}$Rb at 320 K.}
  \label{fig6}
\end{figure}

We also calculate the change of $n_g$ versus $R_{\rm op}$ with the same system
parameters in $^{87}$Rb as the simulation above. The results are shown in
Fig.~\ref{fig6}. The figure shows clearly that as the increment of $R_{\rm op}$
the $n_g$ get more and more smaller and then become negative. The system will
get saturated as the increment of the optical pump rate, which set the
limitation of the largest negative group velocity when $\Omega_{c}$ and
$\Delta$ are fixed. The vertical line in Fig.~\ref{fig6} gives the upper bound
of the optical pump rate derived from Eq.~(\ref{ropeq}) (using
$\Gamma_{52}=\Gamma_{3}/2$), which means the largest
optical pump rate can be got when level $|5\rangle$ is a single energy level. When multiple excited states are used for optical pump or $\Gamma_{52}>\Gamma_{3}/2$, this bound might be exceeded.
Fig.~\ref{fig6} also shows that the Doppler broadening affects only the scale
of $n_g$ and the optical pump strength needed to change the dispersion from
normal to anomalous.

At last we would like to point out that this scheme can be simplified. In a
normal three-level system the Raman peaks can also be flipped by the optical
pump, although the two Raman peaks will not be symmetric. Furthermore, the
normal and anomalous dispersions can also be observed at the root of the Raman
peak when only one coupling is used.

In summary, we have shown a scheme based on the far detuning Raman effect to
control the light speed from slower-than-$c$ to faster-than-$c$. The dispersion
of the system can be controlled by an optical pump, so the velocity of
the probe beam can be changed dynamically by controlling the strength of the
optical pump. Theoretically, this scheme can provide an arbitrary large positive
or negative dispersion, however the largest dispersion is limited by the
bandwidth of the system, which are decided by
the parameters of the system used, e.g. optical depth of the atomic system,
strength of the couplings. 
Our scheme is very easy in realization,
since the requirements are quite simple: a system, which can generate Raman
peaks in far detuning $\Lambda$ structure and an additional level used for
optical pump, can be used to realize this scheme. Almost any system in which
EIT can be observed is suitable to realize this scheme. Although a four-level
system, whose electronic dipole moments fulfill the requirement mentioned in
the second paragraph, will be better, a normal three level structure can be
used too. There is no any technical difficulty too, since the key point of this
scheme is the far detuning Raman peaks which can be flipped using an optical
pump. Fortunately, this has already been demonstrated in
experiment\cite{PRA.73.053804}.

\begin{acknowledgments}
  This work was funded by National Fundamental Research Program (Grant No. 2006CB921907), National Natural Science Foundation of China (Grants No. 60621064,
  No. 10674126, No. 10674127), the Innovation funds from Chinese Academy of Sciences,
  and the program for NCET.
\end{acknowledgments}

% \bibliography{/home/qfeng/papers/papers}

\end{document}